\documentstyle[aps,epsf,psfig]{revtex}
\begin{document}
\draft
\flushbottom

\title{
Spin and superconducting instabilities near a Van Hove singularity}
\author{J. Gonz\'alez \\}
\address{
        Instituto de Estructura de la Materia. 
        Consejo Superior de Investigaciones Cient{\'\i}ficas. 
        Serrano 123, 28006 Madrid. Spain.}
\date{\today}
\maketitle
\widetext
\begin{abstract}
We apply a wilsonian renormalization group approach to 
the system of electrons in a two-dimensional square 
lattice interacting near the saddle-points of the band, 
when the correlations at
momentum ${\bf Q} \equiv (\pi , \pi)$ prevail in the 
system. The detailed consideration of the spin degrees 
of freedom allows to discern the way in which the SU(2)
spin invariance is preserved in the renormalization process.
Regarding the spin correlations, we find two 
different universality classes which correspond, in the 
context of the extended Hubbard model,
to having the bare on-site interaction $U$ repulsive or 
attractive. The first class is characterized by a
spin instability which develops through the condensation
of particle-hole pairs with momentum ${\bf Q}$, with the
disappearance of the Fermi line in the neighborhood of the
saddle-points. Within that class, the attractive or repulsive
character of the nearest-neighbor interaction $V$ dictates
whether there is or not a $d$-wave superconducting instability in
the system. For the Hubbard model with just on-site interaction,
we show that some of the irrelevant operators are able to trigger
the superconducting instability. The naturalness of the competing 
instabilities is guaranteed by the existence of a range of
doping levels in which the chemical potential of the open 
system is renormalized to the level of the saddle-points. 
We incorporate this effect to obtain the phase diagram as a 
function of the bare chemical potential, which displays a 
point of optimal doping separating the regions of 
superconductivity and spin instability.

\end{abstract}
\pacs{             }

\tightenlines

\section{Introduction}

During the last years there has been much effort 
devoted to the study of strongly correlated electron 
systems. The interest has been maintained by the 
behavior displayed by the high-$T_c$ copper-oxide 
compounds since the discovery of their superconductivity 
15 years ago\cite{dago}. There are a number
of features exhibited by these materials that do
not fit into the conventional theoretical frameworks.
The normal state of the cuprates shows for instance
unusual transport properties and, more strikingly, 
a pseudogap phase in which part of the density of 
states is lost at the Fermi level while the system 
remains conducting. It seems that a new paradigm is
needed to describe these materials, in the same way
as the Fermi liquid picture accounts for the behavior
of conventional metals.

From the theoretical point of view, progress has been
made during the past decade in understanding the 
foundations of Landau's Fermi liquid theory and,
consequently, the possible deviations that may open
the way to a new kind of metallic 
behavior\cite{sh,pol,ex}. The most
powerful method used in this task has been the 
renormalization group (RG) approach developed for
interacting fermion systems\cite{sh}. 
We have learned from it
that the Fermi liquid picture is a very robust 
description of the metallic state. There are only a
few perturbations that may destabilize the Fermi 
liquid, favoring the formation of states with 
different types of symmetry breaking. The Fermi liquid 
represents itself a universality class in which any 
electron system falls at dimension 
$D \geq 2$, unless the interaction is 
sufficiently long-ranged\cite{bares,wo,sh2,it,wi,us,ch}
or the Fermi 
surface develops singular points\cite{inst}.

Soon after the discovery of the high-$T_c$ 
superconductivity, it was proposed that the presence of
nonlinear dispersion near the Fermi line of the 
copper-oxide layers could be at the origin of the
unconventional behavior\cite{early,rev}. 
The fermion systems in a
two-dimensional (2D) square lattice have necessarily 
saddle-points in their band dispersion, which 
give rise to Van Hove singularities where the 
density of states diverges logarithmically. In the 
most common instances, the two inequivalent 
saddle-points lie at the boundary of the Brillouin 
Zone, and their hybridization has been proposed to 
explain the existence of a $d$-wave order parameter 
in the superconducting phase\cite{dwave,pin,iof,kohn,liu}, 
as observed experimentally. 
Further investigations have shown that
the unconventional transport properties in the normal
state may be accounted for by the proximity of the 
Fermi level to the Van Hove singularity (VHS) in the
copper-oxide layers\cite{pat,nucl,men,george,kat}.

A careful examination of the kinematics near the 
saddle-points has shown indeed that a superconducting
instability with $d$-wave order parameter arises in
the $t-t'$ Hubbard model with bare repulsive 
interaction\cite{jpn,prl}. 
The mechanism at work is of the same
kind described by Kohn and Luttinger as giving rise
to a $p$-wave pairing instability in the 
three-dimensional Fermi liquid\cite{kl,chu}, but adapted 
now to the 2D model with saddle-points near 
the Fermi line. Other studies have considered in detail
the influence of the entire Fermi line in the 
development of the instabilities of the 
system\cite{ren,ren2}. They 
have given further support to the picture of a 
competition between a spin-density-wave instability
and a pairing instability with $d$-wave order 
parameter in the $t-t'$ Hubbard model with the Fermi
level at the VHS. More recently, a refined  
renormalization program has been implemented in Ref.
\onlinecite{binz} by trying to handle the momentum 
dependence of the vertex functions in the scaling
procedure, what has confirmed the appearance of 
different phases with symmetry breaking in the spin
and the charge sector.

Despite all the results obtained in the system of
electrons near the VHS, there are still important 
obstacles precluding a precise description of the
effective theory at low energies. From a technical
point of view, the source of the problem is the 
appearance of infrared singularities in the RG approach
after accomplishing the renormalization of the leading
logarithm. Some vertex functions, like the four-point
interaction with vanishing total incoming momentum
at the one-loop level or the electron
self-energy at the two-loop level, get $\log^2 
(\Lambda)$ corrections in terms of the energy cutoff
$\Lambda $. After applying the standard RG program,
the renormalized quantities still contain factors of
the form $\log (\Lambda)$. This fact questions the 
predictability of the theory since the argument of
the logarithm has a hidden energy scale, which sets
the strength of the corrections. From a formal point
of view, the theory becomes nonrenormalizable in the 
standard RG approach, since the energy cutoff is not 
the only dimensionful variable that appears in the 
scaling process.

The problem of the renormalizability of the theory can
be best handled by adopting a wilsonian RG approach,
in which only the high-energy modes that live at the
cutoff $\Lambda $ are integrated out at each RG step.
In the present paper we follow Shankar's RG program 
for interacting fermion systems\cite{sh}, which has the 
advantage of decoupling the renormalization of the BCS 
channel (with vanishing momentum of the colliding particles) 
from that of the rest of the channels at the one-loop
level. 

Moreover, the important 
feature of the wilsonian approach is that it allows to
set free the chemical potential, so that it can 
readjust itself at each step of integration. The issue
of the renormalization of the chemical potential has 
been discussed in Ref. \onlinecite{sh} in the context
of Fermi liquid theory, and it reaches great significance
when considering the system of electrons near the VHS.
The chemical potential cannot be fixed at the singularity
from the start, since it is actually the scale needed to 
regularize the infrared singularities that appear in the
standard RG procedure. On the other hand, the final 
location of the chemical potential relative to the VHS 
is not arbitrary, since it is a dynamical quantity that
scales in a predictable way upon renormalization.

We remark that the renormalization devised in the paper   
assumes a constant value of the bare chemical potential, 
instead of a constant particle number of the system.
That is, we describe a
situation appropriate for an electron system in contact
with a charge reservoir, which sets the nominal value
$\mu_0$ of the ensemble. The renormalization accounts for
the reduction suffered by the effective chemical potential
{\em inside} the electron system due to the repulsive
interaction. 
This description of the electron system at constant nominal
chemical potential is most appropriate when dealing with
the Cu-O layers of the cuprate superconductors, since it    
provides a realization of the contact of the 2D layers with
the charge reservoir.
The conclusion is that a variation in the
external chemical potential does not have always a linear 
correspondence with the variation of the final renormalized
value of $\mu $, which is identified with the Fermi energy
of the electron system.

The renormalization of the chemical potential  
makes possible to address the question of the 
naturalness of the picture in which the Fermi level is
fine-tuned to the VHS. 
The strength of the predicted instabilities depends  
crucially on the proximity of the Fermi energy to the
singularity.
This has been the main criticism to the proposals
claiming that the features of the copper-oxide
materials could be related to the properties of
electrons interacting near a VHS. 
We will show that 
the chemical potential is renormalized towards the VHS
in a certain range of filling levels, in such a way that
it may become pinned to the singularity in the
low-energy theory. This fact was already anticipated in 
Refs. \onlinecite{mark}, \onlinecite{newns} 
and \onlinecite{pin}, 
and it has been used to cure the infrared singularities 
of the electron self-energy in Ref. \onlinecite{nucl}. 
In the present paper, we will take into account such an
effect to determine in a predictable way the strength
of the pairing instability in the system, as a function
of the different values of the bare chemical potential.

In the next section we describe the system
to which our analysis applies. In Section III we classify
the different renormalized vertices that arise by explicit
consideration of the spin degrees of freedom. The universality
classes of the system are obtained in Section IV, 
where we also show the way in
which the SU(2) spin invariance is preserved along the
RG flow. Section V is devoted to establish
the properties of the spin instability of the system, while
Section VI analyzes the renormalization of the chemical
potential to determine the region of the phase diagram in
which the superconducting instability prevails. Finally, the
last section is devoted to draw the main conclusions of
this work.

\section{The model}

\begin{figure}
\epsfxsize=5cm 
\centerline{\epsfbox{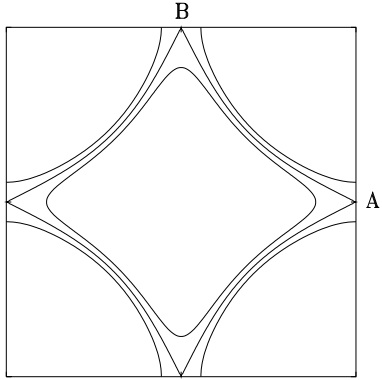}} 
\caption{Contour energy map for the $t-t'$ model
about the Van Hove filling.}
\label{saddle}
\end{figure}

We take as starting point of our analysis a system of 
interacting electrons in the 2D square lattice with 
nearest-neighbor hopping $t$ and next-to-nearest-neighbor
hopping $t'$. The band dispersion of the model is given by
\begin{equation}
\varepsilon ({\bf k}) = -2t (cos (k_x) + cos (k_y))
    + 4t'  cos (k_x) cos (k_y)
\end{equation}
where we have set the lattice spacing equal to one. Some of
the energy contour lines are shown in Fig. \ref{saddle}. The
dispersion has two inequivalent saddle-points $A$ and 
$B$ at the boundary of the Brillouin Zone. In their 
neighborhood, the energy of the one-particle states can be
approximated by the quadratic form
\begin{equation}
\varepsilon_{A,B} ( {\bf k} ) \approx \mp ( t \mp 2 t' ) k_x^2 
\pm ( t \pm 2 t' ) k_y^2 
\label{qua}
\end{equation}
where the momenta $k_x$ and $k_y$ measure now small deviations 
from $A$ and $B$.

As a consequence of the nonlinear character of the dispersion,
the density of states $n(\varepsilon )$ diverges logarithmically
at the level of the saddle-points
\begin{equation}
n(\varepsilon ) \approx  c  \log (t/|\varepsilon|) /(4\pi^2 t)
\label{dos}
\end{equation}
with $c \equiv 1/\sqrt{1 - 4(t'/t)^2}$.
This implies that, when the Fermi level is close to the VHS,
most part of the low-energy states are concentrated in the 
neighborhood of the two saddle-points. In order to apply the 
RG approach, we may take two patches where the quadratic 
approximation (\ref{qua}) holds around the saddle-points.
Higher-order corrections to the expression (\ref{qua}) are
irrelevant under the scaling that makes the action of the model
a fixed-point of the RG transformations, as we see in what 
follows.

We consider then a model whose action at the classical level is
\begin{eqnarray}
S  & = & \sum_{a} \int dt d^2 p \left( i \Psi^+_{a\sigma } 
  ({\bf p}) \partial_t \Psi_{a\sigma } ({\bf p}) 
  - \left( \varepsilon_a ({\bf p}) - \mu_0 \right) 
 \Psi^+_{a\sigma } ({\bf p}) \Psi_{a\sigma } ({\bf p}) 
                               \right)    \nonumber    \\
 &  &  +  \sum_{a,b,c,d} \int dt d^2 p_1 d^2 p_2 d^2 p_3 d^2 p_4
   U ({\bf p}_1, {\bf p}_2, {\bf p}_3, {\bf p}_4) 
  \Psi^+_{a\sigma } ({\bf p}_1)  \Psi^+_{b\sigma'} ({\bf p}_2) 
  \Psi_{c\sigma'} ({\bf p}_4) \Psi_{d\sigma } ({\bf p}_3)
  \delta ( {\bf p}_1 + {\bf p}_2 - {\bf p}_3 - {\bf p}_4 ) 
\label{act}
\end{eqnarray}
where the indices $a,b,c,d$ run over the two patches around 
$A$ and $B$.

The scaling transformation that leaves invariant the kinetic 
term of the action is
\begin{eqnarray}
\partial_t & \rightarrow &  s \partial_t   \label{s1}   \\
{\bf p} & \rightarrow &  s^{1/2} {\bf p}           \\
\Psi_{a\sigma } ({\bf p}) & \rightarrow & 
                       s^{-1/2} \Psi_{a\sigma } ({\bf p})
\label{s3}
\end{eqnarray}
It is easily checked that, with the transformation 
(\ref{s1})-(\ref{s3}), 
the interaction term in the action (\ref{act}) is 
also scale invariant for a constant value of the potential
$U ({\bf p}_1, {\bf p}_2, {\bf p}_3, {\bf p}_4)$. 
If this is not constant, provided that it is a regular 
function of the arguments we can resort to an expansion in
powers of the momenta. Only the constant term is significant,
since the rest of higher-order terms fade away upon scaling
to the low-energy limit $s \rightarrow 0$.
This means that we meet the first
requirement to apply the RG program, that is to have a model
which converges to a fixed-point under RG transformations at 
the classical level.

In the above scaling, we already find the first deviation
in the RG program with respect to the analysis of Fermi liquid
theory. In the case of a model with circular Fermi line, the
interaction term is scale invariant only for very special 
kinematics of the scattering processes\cite{sh}. 
In our model, we have
seen that no constraint is needed on the four momenta involved
in the interaction at the classical level. It is only after 
taking into account virtual processes that the interactions 
will start to grow large under scaling for some particular
choices of the kinematics. This will single out a number 
of so-called marginally relevant channels among all the 
scattering processes, recovering then the similitude with
the analysis of Fermi liquid theory at the quantum level.

The two-patch RG analysis of the $t-t'$ Hubbard model has 
proven to give the dominant instabilities of the system with
the Fermi level at the VHS. For $t' > 0.276 \; t$, a 
ferromagnetic phase has been found below a certain critical
frequency\cite{jpn,prl,kat2,sa}, in agreement with the results 
obtained from Monte Carlo calculations\cite{sor}. 
In this paper we will be interested in
the regime with $t' < 0.276 \; t$, where the competition between
a spin instability and a pairing instability
arises, making the model more appropriate for the comparison
with the phenomenology of the cuprates.

\section{Wilsonian renormalization group}

\begin{figure}
\epsfxsize=7cm 
\centerline{\epsfbox{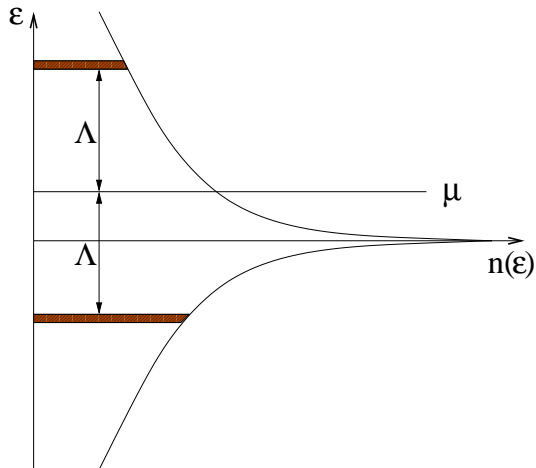}} 
\caption{Picture of the density of states $n(\varepsilon)$
and of the renormalization of the chemical potential $\mu $ 
by integration of states at the energy cutoff $\Lambda $.}
\label{muren}
\end{figure}

In what follows we apply a wilsonian RG approach to 
obtain the low-energy effective theory of the system.
We proceed by progressive integration of the modes in 
two thin shells of width $d \Lambda $ at distance 
$\Lambda $ in energy below and above the Fermi level, as
depicted in Fig. \ref{muren}. For the time being, we
will assume that the Fermi level is located precisely
at the VHS, unless otherwise stated.
This is crucial to obtain
a significant renormalization in any of the interaction
channels, and later on we will comment on the
naturalness of this situation.

The vertex functions may become relevant, that is
increasingly large at low energies, only for very definite
choices of the kinematics.
Focusing on the four-point interaction vertex, this is 
renormalized by a quantity of order $d \Lambda $ at each
RG step only when the momentum transfer along a
pair of external lines is either {\bf 0} or ${\bf Q} \equiv
(\pi ,\pi )$, or when the total momentum of the incoming
modes vanishes (BCS channel). In the present
work we deal with the latter two instances,
since the first corresponds to the case of
forward-scattering interactions, which are subdominant in
the range $t' < 0.276 \; t$ that we are considering.
In this regime, the divergences at vanishing momentum-transfer
are related to charge instabilities of the system, which have
been treated in detail elsewhere\cite{charge}.
We will see that
divergences in the channel with momentum transfer {\bf Q}
give rise to a spin instability, which competes with the
superconducting instability in the BCS channel in the model
with a bare on-site repulsive interaction.

\begin{figure}
\epsfxsize=6.5cm 
\centerline{\epsfbox{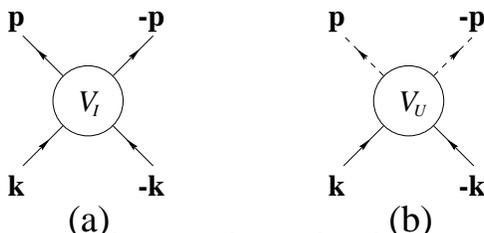}} 
\caption{BCS vertices that undergo renormalization by
particle-particle diagrams. The 
solid and dashed lines stand for modes in the neighborhood 
of the two different saddle points.}
\label{v}
\end{figure}

The different kinematics which may appear in the BCS
channel are listed in Fig. \ref{v}. We allow for the   
possibility of Umklapp processes in which the incoming 
modes scatter from one of the saddle points to the other.

\begin{figure}
\epsfxsize=7cm 
\centerline{\epsfbox{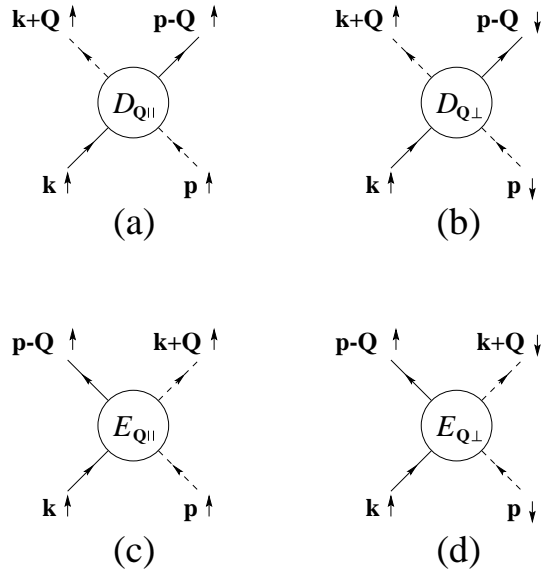}} 
\caption{Direct and exchange vertices that undergo
renormalization by particle-hole diagrams.}
\label{q}
\end{figure}

\begin{figure}
\epsfxsize=7cm 
\centerline{\epsfbox{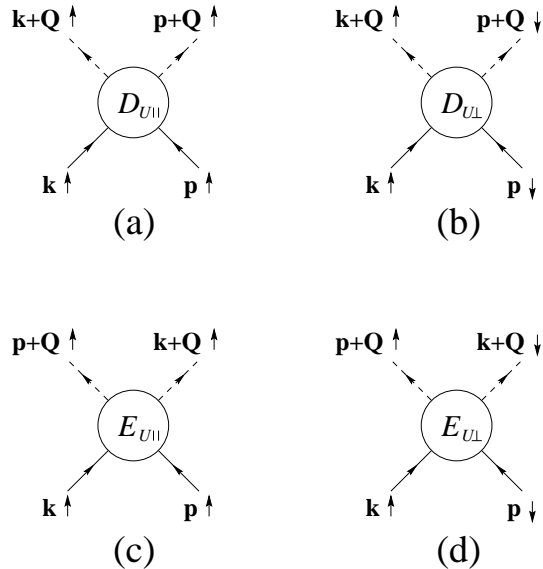}} 
\caption{Umklapp vertices that undergo
renormalization by particle-hole diagrams.}
\label{u}
\end{figure}

The different kinematical possibilities that arise in
the channel with momentum transfer {\bf Q} are classified 
in Figs. \ref{q} and \ref{u}. The first includes the
interactions in which the incoming modes are at different
saddle points, while the latter contains the 
Umklapp processes. The other important distinction is 
between direct ($D$) and exchange ($E$) interactions.
Direct processes are those in which the momentum 
transfer {\bf Q} is taken by the same scattered
fermion line, while in a exchange process the momentum 
transfer takes place between two different fermion lines
connected only by the interaction.

The interaction vertices depicted in Figs. \ref{v}-\ref{u}
are all renormalized upon reduction of the cutoff $\Lambda $.
This can be traced back to the divergent behavior of the
different susceptibilities of the model. By integration of
the high-energy modes in the shells of width $d\Lambda $,
the particle-hole susceptibility at momentum {\bf Q}
gets a contribution 
\begin{equation}
d \chi_{ph} ({\bf Q}) = \frac{c'}{4 \pi^2 t}
            d \Lambda  / \Lambda 
\label{chi1}
\end{equation}
where $c' \equiv \log \left[ \left(1 + \sqrt{1 - 4(t'/t)^2} 
\right)/(2t'/t) \right]$ \cite{lh}.
In the same fashion, the contribution to the
particle-particle susceptibility at zero total momentum is
\begin{equation}
d \chi_{pp} ({\bf 0}) = \frac{c}{4 \pi^2 t} \log (\Lambda )
            d \Lambda  / \Lambda 
\label{chi2}
\end{equation}

In the latter case, the result of the differential integration
diverges logarithmically in the limit $\Lambda \rightarrow 0$.
This has been a source of problems in the usual RG analyses of
the model. The definition of the argument in the logarithm
needs an additional scale, while a proper RG scaling requires
that the energy is the only dimensionful variable in the
problem. It has to be realized that the coefficient at the
right-hand-side of Eq. (\ref{chi2})
represents actually the density of states. This has to be
born in mind for the correct implementation of the RG
approach, as we will discuss later.

\begin{figure}
\begin{center}
\mbox{\epsfxsize 6cm \epsfbox{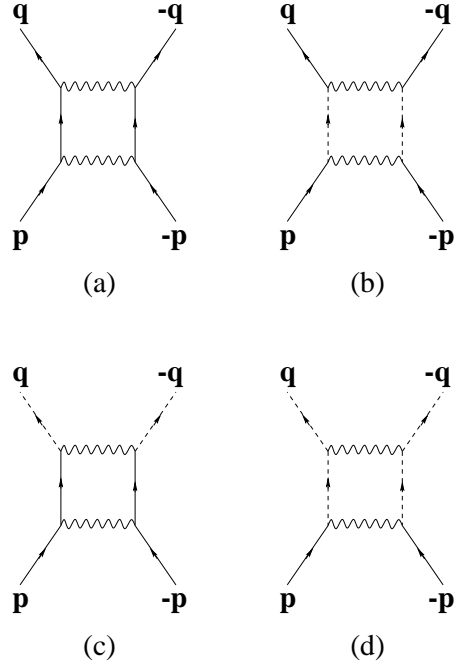}}
\end{center}
\caption{Particle-particle diagrams renormalizing the BCS
vertices at the one-loop level.}
\label{rintra2}
\end{figure}

\begin{figure}
\begin{center}
\mbox{\epsfxsize 5cm \epsfbox{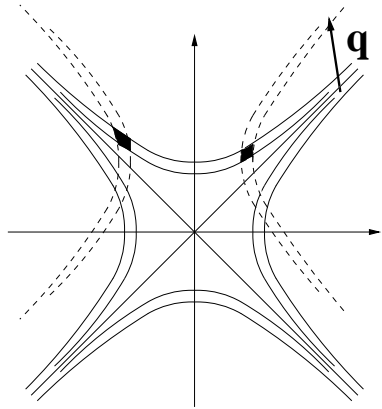}}
\end{center}
\caption{Picture of the high-energy shells of width $d\Lambda $
at a given saddle-point. The dark regions represent the
contribution to a particle-hole diagram when ${\bf q}$ is the 
total incoming momentum.}
\label{ph}
\end{figure}

Let us deal first with the renormalization of the vertices
with BCS kinematics in Fig. \ref{v}. At the one-loop level,
the vertices $V_{I}$ and $V_{U}$ get corrections of
order $d \Lambda /\Lambda $ from the diagrams shown in
Fig. \ref{rintra2}. It is
important to realize that these are the only diagrams
to be taken into account to first order in $d \Lambda $.
There are also corrections from particle-hole diagrams
but, as long as the momentum that goes into the particle-hole
loop is not precisely zero or {\bf Q}, these terms are of 
order $(d \Lambda)^2 $ and therefore irrelevant in the
low-energy limit, as shown graphically in Fig. \ref{ph}.

The BCS vertices mix between themselves alone at the one-loop
level, and the situation is similar in that respect to the
general analysis of the 2D Fermi liquid\cite{sh}. The degree of
renormalization depends on the density of states
$n(\varepsilon )$ at the shells integrated out. For later
use, we consider at this point the most general case in which 
the chemical potential $\mu $ does not coincide from the start 
with the level of the VHS. The differential RG equations take then
the form
\begin{eqnarray}
\Lambda \frac{\partial V_{I}}{\partial \Lambda}
   & = &    c \; n(\mu - \Lambda ) \;
     \left( V_{I}^2 + V_{U}^2 \right)   \label{iu1}  \\
\Lambda \frac{\partial V_{U}}{\partial \Lambda}
   & = &   2  c \; n(\mu - \Lambda ) \;
     V_{I} V_{U}
\label{iu2}
\end{eqnarray}
These equations were
considered in Ref. \onlinecite{jpn}, and they also appear 
as the leading order in the RG approach of Ref.
\onlinecite{binz}.

\begin{figure}
\begin{center}
\mbox{\epsfxsize 6cm \epsfbox{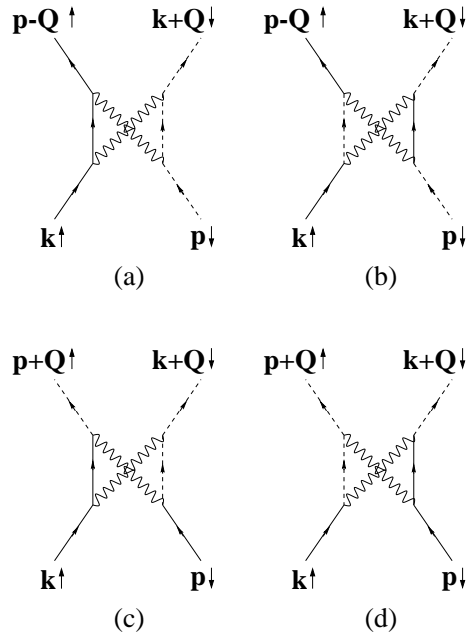}}
\end{center} 
\caption{Particle-hole diagrams renormalizing the vertices
$E_{{\bf Q}\perp }$ and $E_{U\perp }$ at the one-loop level.}
\label{rexch}
\end{figure}

\begin{figure}
\begin{center}
\mbox{\epsfxsize 5cm \epsfbox{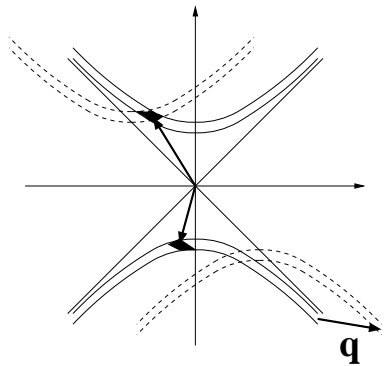}}
\end{center}
\caption{Same scheme as in Fig. \ref{ph}. The dark regions 
represent the contribution to a particle-particle diagram when 
${\bf q}$ is the total incoming momentum.}
\label{pp}
\end{figure}

We consider next the renormalization of the vertices 
$E_{{\bf Q} \perp}$ and $E_{U \perp}$, which have also the
property that they mix only between themselves in the one-loop
corrections linear in $d \Lambda $. These have been represented
in Fig. \ref{rexch}. It can be checked
that any other diagrams give irrelevant contributions of order
$(d \Lambda )^2$, because they involve either a particle-hole
susceptibility at momentum different from {\bf Q} or a
particle-particle susceptibility with total momentum different
from zero. In the latter case, for instance, it is shown in Fig.
\ref{pp} that the number of intermediate states produced by
integration of high-energy modes is quadratic, instead of
linear in $d \Lambda $.

The differential RG equations for the pair of vertices read
\begin{eqnarray}
\Lambda \frac{\partial E_{{\bf Q}\perp }}{\partial \Lambda}
 &  = &   - c' ( E_{{\bf Q}\perp }^2 + E_{U\perp }^2 )
                              /(4 \pi^2 t)     \label{exch1} \\
\Lambda \frac{\partial  E_{U\perp } }{\partial \Lambda}
 &  = &   - c' E_{{\bf Q}\perp }  E_{U\perp }  /(2 \pi^2 t)  
\label{exch2}
\end{eqnarray}
These equations were obtained in Ref. \onlinecite{jpn}, where
the names $U_{inter}$ and $U_{umk}$ were used instead of
$E_{{\bf Q} \perp}$ and $E_{U \perp}$ introduced in the present
paper. The same
equations also arise at the dominant level in the functional
renormalization of Ref. \onlinecite{binz}.

We now turn to the rest of the vertices, $D_{{\bf Q}\parallel }$,
$D_{{\bf Q}\perp }$, $E_{{\bf Q}\parallel }$, $D_{U\parallel }$,
$D_{U\perp }$ and $E_{U\parallel }$, which renormalize among
themselves at the one-loop level. It is clear that 
the vertices $D_{{\bf Q}\parallel }$ and $E_{{\bf Q}\parallel }$
cannot be distinguished from each other just by looking at
the external legs. The same applies to $D_{U\parallel }$
and $E_{U\parallel }$. At the one-loop level, one can still
discern whether the momentum transfer {\bf Q} takes place along
the same scattered fermion line or not. However, the different
corrections have to organize so that the above pairs of
vertices enter in the combinations $D_{{\bf Q}\parallel } -
E_{{\bf Q}\parallel }$ and $D_{U\parallel } - E_{U\parallel }$,
which are the quantities that make physical sense. In that
respect, the situation is similar to what happens with the
couplings $g_{1\parallel }$ and $g_{2\parallel }$ in the
one-dimensional electron systems\cite{lutt1}.

\begin{figure}  
\begin{center}
\mbox{\epsfxsize 7cm \epsfbox{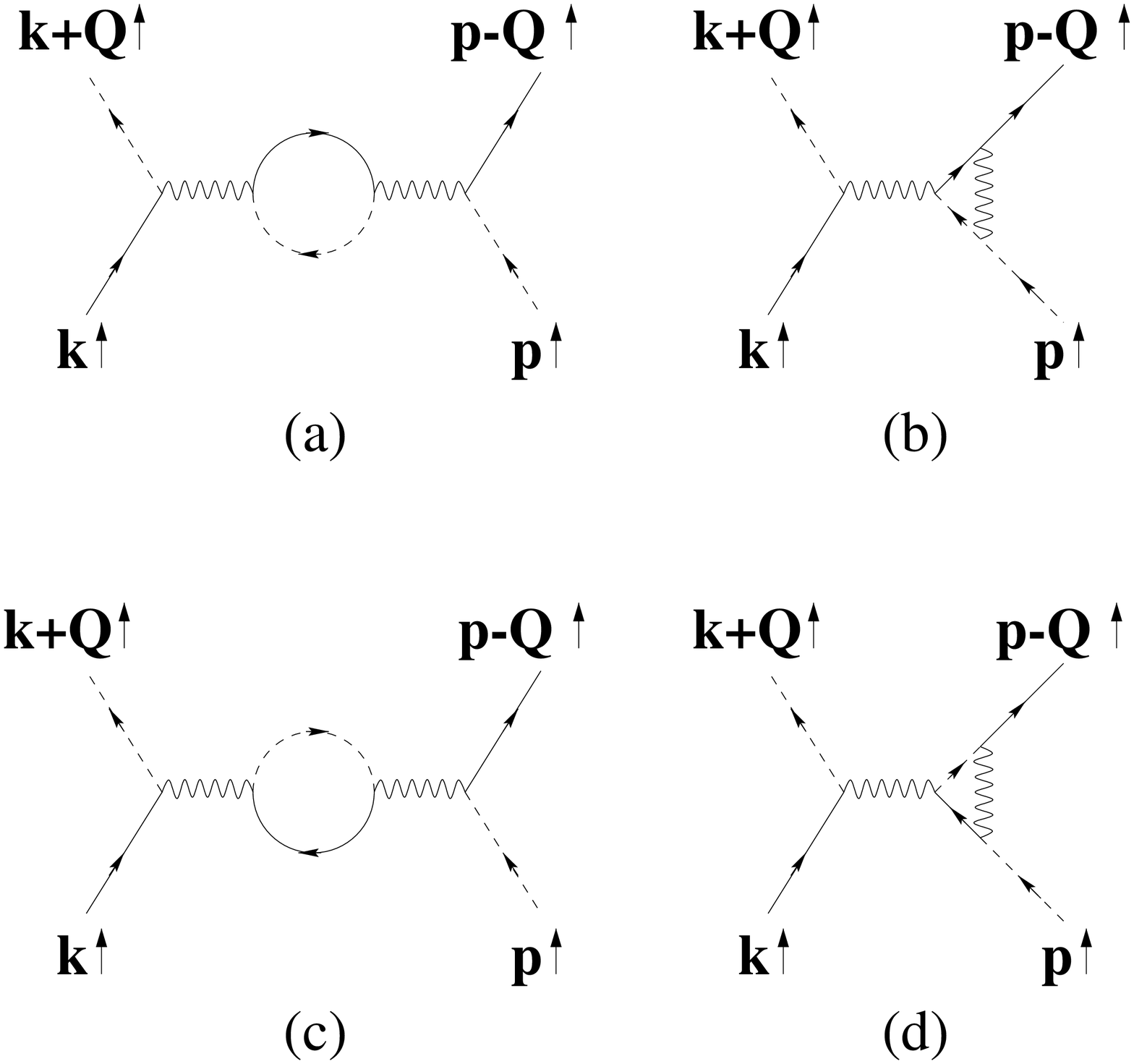}}
\end{center}
\caption{Particle-hole diagrams renormalizing the vertex
$D_{{\bf Q}\parallel }$ at the one-loop level.}
\label{rdir}
\end{figure}

\begin{figure}
\begin{center}
\mbox{\epsfxsize 6cm \epsfbox{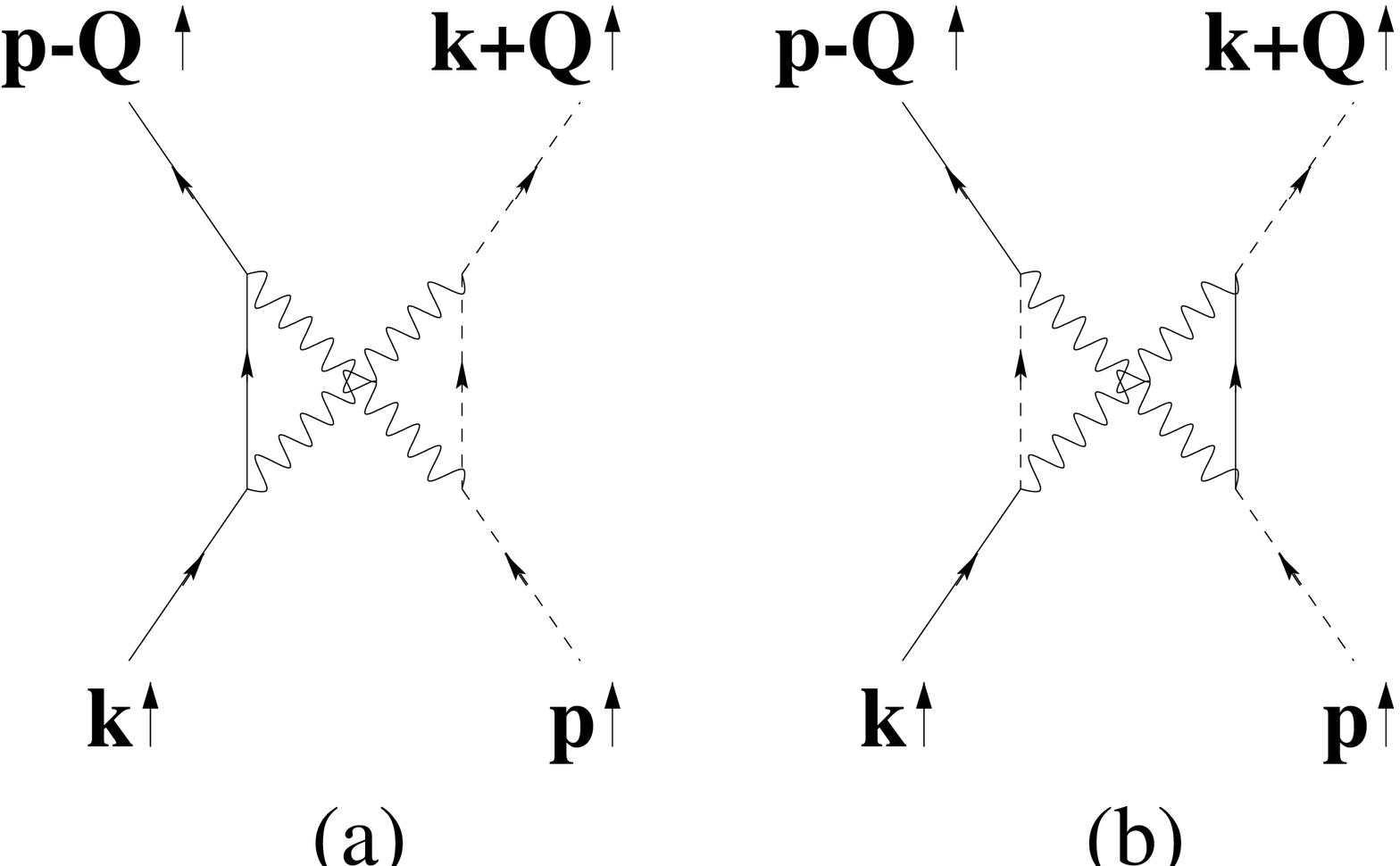}}
\end{center} 
\caption{Particle-hole diagrams renormalizing the vertex
$E_{{\bf Q}\parallel }$ at the one-loop level.}
\label{rexchp}
\end{figure}

The one-loop renormalization of the vertices provides
an explicit proof of the above statement. The vertex
$D_{{\bf Q}\parallel }$ gets linear corrections in $d\Lambda $
from the diagrams shown in Fig. \ref{rdir}, while
$E_{{\bf Q}\parallel }$ is renormalized by the diagrams
shown in Fig. \ref{rexchp}. Their RG equations read then
\begin{eqnarray}
\Lambda   \frac{\partial D_{{\bf Q} \parallel} }
       {\partial \Lambda}    & = &
  c' \left(  D_{{\bf Q} \parallel }^2 + D_{{\bf Q} \perp }^2
   + D_{U \parallel }^2 + D_{U \perp }^2
   - 2 D_{{\bf Q} \parallel } E_{{\bf Q} \parallel}
   - 2 D_{U \parallel } E_{U \parallel}   \right)
                           /(4 \pi^2 t)    \label{fdqpa}   \\
\Lambda   \frac{\partial E_{{\bf Q} \parallel} }
       {\partial \Lambda}    & = &
 - c'  \left(  E_{{\bf Q} \parallel }^2 + E_{U \parallel }^2
              \right)  /(4 \pi^2 t)                    \\
\end{eqnarray}
These two equations can be combined to be written in terms of
the physical vertex,
\begin{equation}
\Lambda   \frac{\partial \left( D_{{\bf Q} \parallel}
                 - E_{{\bf Q} \parallel} \right)  }
       {\partial \Lambda}     = 
    c'  \left[ \left( D_{{\bf Q} \parallel} -
                         E_{{\bf Q} \parallel} \right)^2
     + \left( D_{U \parallel} - E_{U \parallel} \right)^2
         + D_{{\bf Q} \perp }^2 + D_{U \perp }^2   \right]
                           /(4 \pi^2 t)              
\end{equation}

The RG equations for the remaining vertices also depend on the
combinations $D_{{\bf Q}\parallel } - E_{{\bf Q}\parallel }$
and $D_{U\parallel } - E_{U\parallel }$. In the case of 
$D_{{\bf Q}\perp }$, we have 
\begin{equation}
\Lambda   \frac{\partial  D_{{\bf Q} \perp }  }
       {\partial \Lambda}     = 
    c'  \left[ \left( D_{{\bf Q} \parallel} -
       E_{{\bf Q} \parallel} \right)  D_{{\bf Q} \perp }
    + \left( D_{U \parallel} - E_{U \parallel} \right)
                       D_{U \perp }         \right]
                           /(2 \pi^2 t)
\label{fdqpe}
\end{equation}

Finally, the RG equations for $D_{U\parallel }$, $D_{U\perp }$ and
$E_{U\parallel }$ take the form
\begin{eqnarray}
\Lambda   \frac{\partial  D_{U \parallel }  }
       {\partial \Lambda}    & = &
     c' \left(   D_{{\bf Q} \parallel}  D_{U \parallel }
         -    D_{{\bf Q} \parallel}  E_{U \parallel }
         -    D_{U \parallel}  E_{{\bf Q} \parallel }
         +    D_{{\bf Q} \perp }  D_{U \perp } \right)
                           /(2 \pi^2 t)                 \\
\Lambda   \frac{\partial  E_{U \parallel }  }
       {\partial \Lambda}    & = &
    -  c'   E_{{\bf Q} \parallel}  E_{U \parallel }
                           /(2 \pi^2 t)                 \\
\Lambda   \frac{\partial  D_{U \perp }  }
       {\partial \Lambda}    & = &
     c'  \left[ \left( D_{{\bf Q} \parallel} -
       E_{{\bf Q} \parallel} \right)  D_{U \perp }
    + \left( D_{U \parallel} - E_{U \parallel} \right)
                       D_{{\bf Q} \perp }         \right]
                           /(2 \pi^2 t)
\end{eqnarray}
As a final check, the equation for $D_{U\parallel }
- E_{U\parallel }$ turns out to depend on the physical
combination of couplings
\begin{equation}
\Lambda   \frac{\partial \left( D_{U \parallel}
    - E_{U \parallel} \right)  }{\partial \Lambda}  = 
    c'  \left[ \left( D_{{\bf Q} \parallel} -
       E_{{\bf Q} \parallel} \right)
     \left( D_{U \parallel } - E_{U \parallel }   \right)
    +   D_{{\bf Q} \perp }  D_{U \perp }    \right]
                           /(2 \pi^2 t)
\end{equation}

\section{Universality classes}

We discuss now the universality classes in which
the system may fall regarding the spin correlations.
We will focus on the analysis of bare repulsive interactions,
that is where the competition between spin and superconducting
instabilities arises. We will see that our RG scheme is able
to preserve the spin-rotational invariance of models whose
bare interactions have such a symmetry. This provides 
another nontrivial check of our RG approach, as our
framework offers the possibility to analyze the scaling of
interactions with and without the SU(2) spin symmetry.

The interactions of physical interest have the property that
$D_{{\bf Q}\parallel } - E_{{\bf Q}\parallel } =
D_{U\parallel } - E_{U\parallel }$ and $D_{{\bf Q} \perp } =
D_{U \perp }$. These conditions are maintained along
the RG flow if they are satisfied by the bare couplings. Thus,
it is useful to work with the set of couplings
\begin{eqnarray}
D^{\pm }_{\parallel} & \equiv  &
   D_{{\bf Q}\parallel } - E_{{\bf Q}\parallel } \pm
   D_{U\parallel } - E_{U\parallel }                    \\
D^{\pm }_{\perp } & \equiv  &
        D_{{\bf Q} \perp } \pm  D_{U \perp }
\end{eqnarray}
From the results of the preceding section, these new couplings
satisfy the equations
\begin{eqnarray}
\Lambda   \frac{\partial  D^{\pm }_{\parallel }  }
       {\partial \Lambda}    & = &
    c'  \left[ (  D^{\pm }_{\parallel}  )^2
      +   (  D^{\pm }_{\perp }  )^2    \right]
                           /(4 \pi^2 t)   \label{dpml}   \\
\Lambda   \frac{\partial  D^{\pm }_{\perp }  }
       {\partial \Lambda}    & = &
    c'   D^{\pm }_{\parallel}   D^{\pm }_{\perp }
                           /(2 \pi^2 t)    \label{dpmp}   
\end{eqnarray}

\begin{figure}
\epsfxsize=7cm 
\centerline{\epsfbox{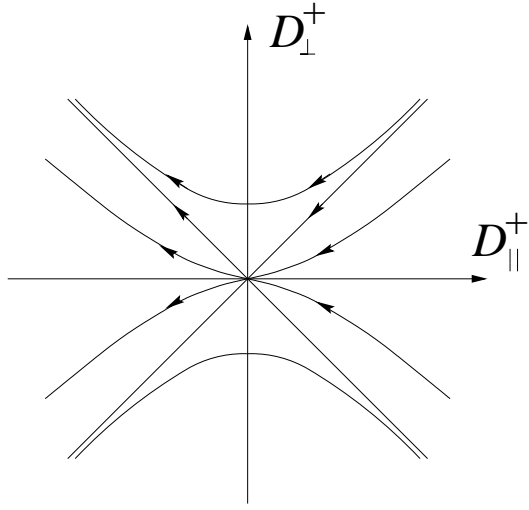}}
\caption{Flow of the renormalized couplings in the
$( D^{+}_{\parallel } , D^{+}_{\perp }  )$
plane.}
\label{flowdd}
\end{figure}

The universality classes of the system 
can be obtained from the integrals of Eqs. (\ref{dpml})
and (\ref{dpmp}). We stick to the case
in which $D^{-}_{\parallel } = D^{-}_{\perp } = 0$. The
flow for the couplings $D^{+}_{\parallel }$ and $D^{+}_{\perp }$
is represented in Fig. \ref{flowdd}. Focusing on interactions
that are repulsive at the initial stage of the RG, that is
$D^{+}_{\parallel } > 0$ and $D^{+}_{\perp } > 0$, we observe
two possible behaviors of the renormalized couplings. In
the case in which the bare couplings satisfy $D^{+}_{\parallel }
\geq D^{+}_{\perp }$, the flow is bounded and it converges
monotonically to the origin of the space of couplings. If
we start otherwise from a point with $D^{+}_{\parallel } <
D^{+}_{\perp }$, the flow becomes unstable and it approaches
a regime in which $D^{+}_{\parallel } \rightarrow -\infty $
and $D^{+}_{\perp } \rightarrow  +\infty $.

The regions with stable and unstable flow correspond to
respective universality classes, which imply quite different
physical properties. Let us focus, for instance, on the extended
Hubbard model with on-site interaction $U$ and interaction $V$
between nearest-neighbor sites. The appropriate bare values
for the couplings in Figs. \ref{q} and \ref{u} are
\begin{eqnarray}
 D_{{\bf Q}\parallel } = D_{{\bf Q}\perp } & = & U - 4 V
                       \label{first}            \\
 E_{{\bf Q}\parallel } = E_{{\bf Q}\perp } & = & U + \alpha V \\
   D_{U\parallel } = D_{U\perp } & = & U - 4 V      \\
   E_{U\parallel } = E_{U\perp } & = & U - \beta V
\label{last}     
\end{eqnarray}
with $0 < \alpha, \beta < 4$. We have for the initial values
of the flow $D^{+}_{\parallel } = -(8 + \alpha - \beta ) V$ and
$D^{+}_{\perp } = 2 U - 8 V$. With the physically sensible choice
$\alpha = \beta $, we see that the attractive or repulsive
character of the on-site interaction dictates whether the
RG flow is bounded or not in the upper half-plane of
Fig. \ref{flowdd}.

The fact that the flow is not bounded for $U > 0$
points to the development of some instability in the system.
The divergence of the renormalized couplings represents the 
failure to describe the model in terms of the original
fermion variables. The underlying physical effect is the
condensation of boson degrees of freedom, as we will show
in the next section. The preservation of the spin-rotational
invariance at each step of the RG process helps to clarify
the physical interpretation of the instability and to discern
the issue of the spontaneous breakdown of the symmetry.

We pay attention then to the way in which the SU(2) spin
symmetry is preserved in our RG framework. This can be analyzed
by looking at the response functions for the different
components of the spin operator. Since the renormalized
interactions grow large at momentum transfer ${\bf Q} =
(\pi ,\pi )$, we focus on the correlations of the operator
\begin{equation}
 S_j ({\bf Q}) = \sum_{k} 
\Psi^{+}_{\sigma} ({\bf k} + {\bf Q})
       \sigma^{\sigma \sigma '}_j
     \Psi_{\sigma '} ({\bf k}) \;\;\;\;\;\;\; j = x,y,z
\label{spinq}
\end{equation}

\begin{figure}
\begin{center}
\mbox{\epsfxsize 8cm \epsfbox{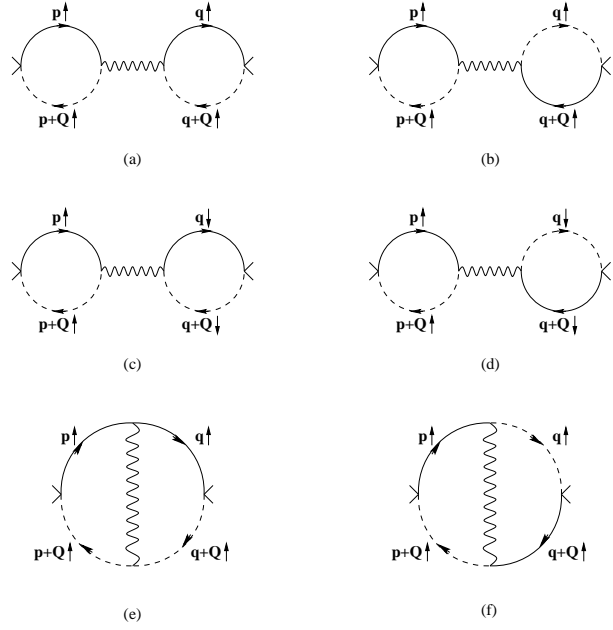}}
\end{center}
\caption{First-order contributions to the correlator of the
$S_z$ operator.}
\label{rz}               
\end{figure}

The scaling properties of the response functions can be studied
in the same fashion as for the interacting one-dimensional
fermion systems\cite{sch}. 
The response function $R_z (\omega )$ for
the $S_z ({\bf Q}) $ operator, for instance, is renormalized
by the diagrams shown in Fig. \ref{rz}. After taking the
derivative with respect to the cutoff and imposing the
self-consistency of the diagrammatic expansion, we obtain
\begin{equation}
\frac{\partial R_{z}}{\partial \Lambda} =
   - \frac{2c'}{\pi^2 t}    \frac{1}{\Lambda}  +
     \frac{c'}{ \pi^2 t} \left( D_{{\bf Q}\parallel } -
  E_{{\bf Q}\parallel } + D_{U\parallel } - E_{U\parallel }
     - D_{{\bf Q}\perp } - D_{U\perp }   \right)
                 \frac{1}{\Lambda}  R_{z}
\label{z}
\end{equation}

\begin{figure}
\epsfxsize=8cm 
\centerline{\epsfbox{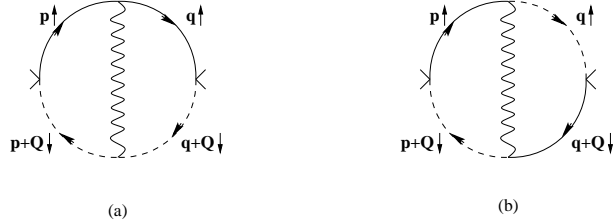}} 
\caption{First-order contributions to the correlators
of the $S_x$ and $S_y$ operators.}
\label{rx}
\end{figure}

The response functions $R_x (\omega )$ and $R_y (\omega )$ for
the other two components of the spin operator are both
renormalized by the diagrams shown in Fig. \ref{rx}. Following
the same procedure as for $R_z (\omega )$, we obtain
\begin{equation}
\frac{\partial R_{x}}{\partial \Lambda} =
   - \frac{2c'}{\pi^2 t}    \frac{1}{\Lambda}  -
     \frac{c'}{ \pi^2 t} \left(
      E_{{\bf Q}\perp } + E_{U\perp }   \right)
                 \frac{1}{\Lambda}  R_{x}
\label{x}
\end{equation} 
and a completely similar equation for $R_y (\omega )$.

The response functions $R_x (\omega ), R_y (\omega )$ and
$R_z (\omega )$ can be made exactly equal if the equation
\begin{equation}
D_{{\bf Q}\parallel } - E_{{\bf Q}\parallel }
         + D_{U\parallel } - E_{U\parallel }  
 - D_{{\bf Q}\perp } - D_{U\perp }    =
 - E_{{\bf Q}\perp } - E_{U\perp }  
\end{equation}
is satisfied all along the
flow. From Eqs. (\ref{exch1}), (\ref{exch2}), (\ref{dpml}), and
(\ref{dpmp}), we observe that this is automatically fulfilled
when the condition is imposed for the initial values of the
couplings. In the case of the extended Hubbard model, we have
indeed for the bare couplings in Eqs. (\ref{first})-(\ref{last})
\begin{equation}
D_{{\bf Q}\perp } + D_{U\perp }
 - D_{{\bf Q}\parallel } + E_{{\bf Q}\parallel }
         - D_{U\parallel } + E_{U\parallel }  =
   E_{{\bf Q}\perp } + E_{U\perp }  =  2U + (\alpha - \beta )V
\end{equation}
The condition
is actually satisfied by the couplings of any hamiltonian
that is invariant under rotations. We show in this way that
the SU(2) spin symmetry can be preserved at each point of the
RG flow of the couplings, so that the low-energy effective
action keeps the invariance of the bare hamiltonian.

\section{Spin instability}

We proceed to determine the physical properties of the
universality class corresponding to the unstable flow in
the upper half-plane of Fig. \ref{flowdd}. The divergence
of the renormalized couplings $D^{+}_{\perp } -
D^{+}_{\parallel }$ and $E_{{\bf Q}\perp } + E_{U\perp }$
results in the divergence of the response functions 
$R_x$, $R_y$ and $R_z$ at a certain value of their argument.
This points at the development of an instability in the spin
sector at the corresponding value of the energy measured
from the Fermi level.

The divergence of the response functions implies the 
existence of a pole at a given frequency $\omega_c $.
From the solution to Eqs. (\ref{exch1}), (\ref{exch2}), 
(\ref{dpml}), and (\ref{dpmp}), the value of the pole 
is given by
\begin{equation}
1 - ( D^{+}_{\perp }(\Lambda_0 )
        -  D^{+}_{\parallel }(\Lambda_0 ) )
            \chi_{ph} ({\bf Q}, \omega_c ) = 0
\label{pole}
\end{equation}
where $D^{+}_{\perp }(\Lambda_0 )$ and
$D^{+}_{\parallel }(\Lambda_0 )$ are the initial values of
the couplings.
As long as the susceptibility $\chi_{ph} $ at momentum
${\bf Q}$ diverges logarithmically in the low-frequency
limit, it is clear that the above condition is satisfied
no matter how small the initial value of the coupling
$D^{+}_{\perp } - D^{+}_{\parallel }$ may be.

It is important to bear in mind that the 
susceptibility $\chi_{ph} $ at momentum ${\bf Q}$ has 
a finite imaginary part, which is essential to discern
the nature of the ground state of the system. The 
imaginary part is computed in the Appendix, and it
turns out to be $ c' /(8 \pi t) $. The equation
(\ref{pole}) can be written then in the form
\begin{equation}
1 - ( D^{+}_{\perp }(\Lambda_0 )
      -  D^{+}_{\parallel }(\Lambda_0 ) )
 \frac{c'}{4 \pi^2 t}  \log(i\Lambda_0 /\omega_c ) = 0
\end{equation}
which shows that the pole occurs for a pure imaginary 
value $\omega_c = i |\omega_c |$.

The appearance of a pole in the correlator of a boson
operator for a pure imaginary frequency corresponds to a
phenomenon of condensation, in the same fashion as it
happens in the case of a pairing instability\cite{agd}.
In the present instance, the boson-like object is 
the spin operator at momentum ${\bf Q}$ defined in Eq.
(\ref{spinq}). The fact that the pole arises at a value
$i |\omega_c |$ means that the instability pertains
actually to the theory posed at finite temperature, 
and that there is a transition to a condensed phase
at a temperature of the order of magnitude given by
$|\omega_c |$.

In our case, the boson operator that acquires a 
nonvanishing mean value due to the spin instability
is the vector $ \int d^2 k d\omega \Psi^{+}_{\sigma} ({\bf k}) 
\mbox{\boldmath $\sigma $}^{\sigma \sigma' } \Psi_{\sigma'}
({\bf k}+{\bf Q}) $.
This has important consequences, since the diagrammatic
approach has to be rebuilt below the point of the 
transition, in the same way as in the 
case of a pairing instability\cite{landau}.

Let us focus on the Hubbard model, i. e. on a model with
interaction between currents with opposite spin
projections. To fix ideas,
suppose that the vector \mbox{\boldmath $S$} gets the nonzero
mean value pointing in the $x$ direction. Then, there
are two different kinds of one-particle propagators, since
the presence of the condensate leads to the consideration
of correlators of the type $\langle
\Psi^{+}_{A \uparrow } ({\bf k}, \omega)
     \Psi_{B \downarrow} ({\bf k}, \omega) \rangle$,
as well as of the usual propagators for well-defined spin
projection near each of the saddle-points. To include all
the different possibilities, we define the propagator
$G_{a\sigma, b\sigma' }({\bf k}, \omega )$, with indices
$a,b$ labelling the saddle-points and $\sigma,\sigma'$
labelling the spin projections:
\begin{equation}
G_{a\sigma, b\sigma' }({\bf k}, \omega ) =
i  \langle  \Psi^{+}_{a \sigma } ({\bf k}, \omega)
     \Psi_{b \sigma' } ({\bf k}, \omega) \rangle
\end{equation}

\begin{figure}
\epsfxsize=8.5cm 
\centerline{\epsfbox{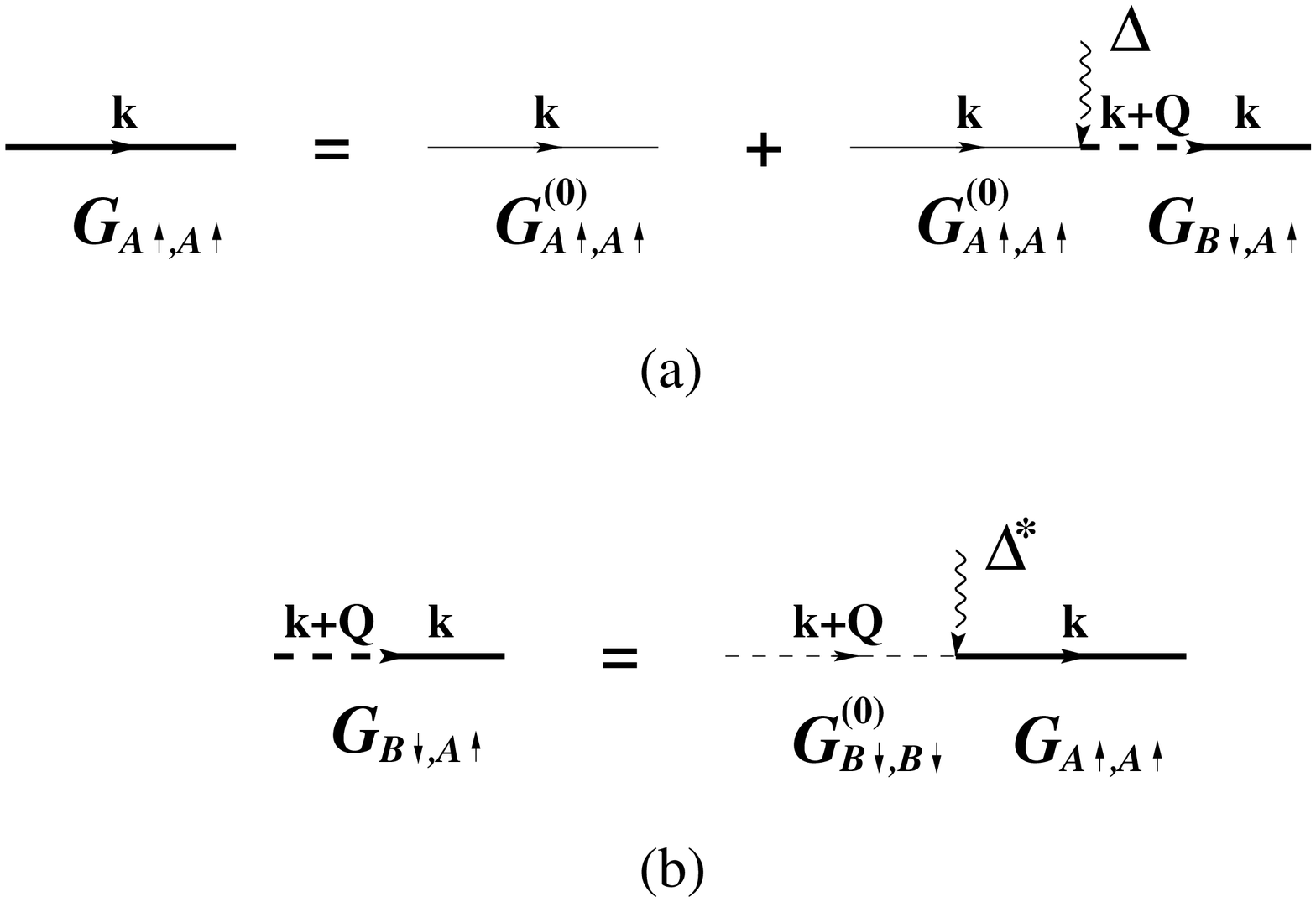}} 
\caption{Self-consistent equations for the dressed propagators
in the particle-hole condensate, in terms of the undressed
propagators at the two inequivalent saddle-points.}
\label{schw}
\end{figure}

The Schwinger-Dyson equations for the one-particle
propagators take the form shown graphically in Fig. \ref{schw},
where the insertion of the wavy line represents the
factor
\begin{equation}
U \int d^2 k d\omega \langle \Psi^{+}_{A\uparrow} ({\bf k})
 \Psi_{B\downarrow} ({\bf k}+{\bf Q}) \rangle  \equiv \Delta
\label{gap}
\end{equation}
We have, for instance, the closed set of equations
\begin{eqnarray}
G_{A\uparrow,A\uparrow} & = & G^{(0)}_{A\uparrow,A\uparrow}
 + G^{(0)}_{A\uparrow,A\uparrow} \Delta
                  G_{B\downarrow,A\uparrow}   \label{schw1}    \\
G_{B\downarrow,A\uparrow} & = & G^{(0)}_{B\downarrow,B\downarrow}
   \Delta^{*}  G_{A\uparrow,A\uparrow}        \label{schw2}
\end{eqnarray}
where the superindex $0$ denotes the corresponding propagator
before the introduction of the condensate.
Eqs. (\ref{schw1}) and (\ref{schw2}) can be combined
to give an equation for $G_{A\uparrow,A\uparrow}$, which reads
\begin{equation}
G_{A\uparrow,A\uparrow}  =  G^{(0)}_{A\uparrow,A\uparrow}
 + G^{(0)}_{A\uparrow,A\uparrow} \Delta 
  G^{(0)}_{B\downarrow,B\downarrow}  \Delta^{*}  
               G_{A\uparrow,A\uparrow}
\label{propa}
\end{equation}

The solution to Eq. (\ref{propa}) takes the form
\begin{equation}
G_{A\uparrow,A\uparrow} ({\bf k},\omega)  = 
 \frac{ G^{(0)}_{A\uparrow,A\uparrow} ({\bf k},\omega) }
 { 1 - G^{(0)}_{A\uparrow,A\uparrow} ({\bf k},\omega) |\Delta|^2
  G^{(0)}_{B\downarrow,B\downarrow} ({\bf k},\omega)  } 
\label{sol}
\end{equation}
in terms of the propagators at the two different saddle-points
\begin{eqnarray}
G^{(0)}_{A\uparrow,A\uparrow} ({\bf k},\omega) & = & \frac{1}
          {\omega  - \varepsilon_A ({\bf k}) 
     + i \epsilon \; {\rm sgn}  (\omega )  } \\
G^{(0)}_{B\downarrow,B\downarrow} ({\bf k},\omega) & = & \frac{1}
    {\omega  - \varepsilon_B ({\bf k}) 
          + i \epsilon \; {\rm sgn} (\omega ) }
\end{eqnarray}
The important point is to determine the pole structure of the 
propagator (\ref{sol}). Its frequency dependence can be 
expressed in the form 
\begin{eqnarray}
G_{A\uparrow,A\uparrow} ({\bf k},\omega)  & = & 
\frac{\omega  - \varepsilon_B ({\bf k})}
    {\left(\omega  - \varepsilon_A ({\bf k}) + 
      i \epsilon \; {\rm sgn} (\omega )   \right) 
      \left(\omega  - \varepsilon_B ({\bf k})  +
    i \epsilon \; {\rm sgn} (\omega ) \right)   
                 - |\Delta|^2  }          \label{dec1}      \\
   & = &  \frac{u({\bf k})^2}{\omega - \varepsilon_u ({\bf k})
                  +   i \epsilon \; {\rm sgn} (\omega ) }  +
    \frac{v({\bf k})^2}{\omega - \varepsilon_v ({\bf k})
                  +   i \epsilon \; {\rm sgn} (\omega ) }
\label{dec2}
\end{eqnarray}
with appropriate weights $u({\bf k})^2 , v({\bf k})^2$, and
$\varepsilon_u ({\bf k}) , \varepsilon_v ({\bf k})$ being the 
roots of the denominator in Eq. (\ref{dec1})
\begin{equation}
\varepsilon_{u,v} ({\bf k}) = \left( \varepsilon_A ({\bf k}) +
  \varepsilon_B ({\bf k})  \pm \sqrt{ (\varepsilon_A ({\bf k}) 
     -\varepsilon_B ({\bf k}) )^2 + 4 |\Delta|^2 } \right) / 2
\end{equation}

From the physical point of view, the most important feature is
the appearance of a gap in the quasiparticle spectrum near the
saddle-points. This can be checked by determining the shape 
of the Fermi line, which is given by setting either $\varepsilon_u 
({\bf k}) = 0$ or $\varepsilon_v ({\bf k}) = 0$. Both conditions
lead to the equation 
\begin{equation}
\varepsilon_A ({\bf k}) \varepsilon_B ({\bf k}) - |\Delta|^2 = 0
\end{equation}
By recalling that
$\varepsilon_A ({\bf k}) = - t_{-}k_x^2 + t_{+}k_y^2$ and 
$\varepsilon_B ({\bf k}) = t_{+}k_x^2 - t_{-}k_y^2$, 
we end up with the equation satisfied by the points of the 
Fermi line
\begin{equation}
(t_{-}k_x^2 - t_{+}k_y^2) (t_{+}k_x^2 - t_{-}k_y^2 )
 + |\Delta|^2 = 0
\label{fl}
\end{equation}

Solving Eq. (\ref{fl}) for the variable $k_y^2$, for instance, 
we find that there is a solution only for values of $k_x^2$  
such that
\begin{equation}
(t_{+}^2 - t_{-}^2)^2 k_x^4 - 4 |\Delta|^2 t_{+} t_{-} \geq 0
\end{equation}
Reminding that $t_{\pm} \approx t \pm 2t'$, this condition
implies that, for small values of $t'$, there is a gap in the 
spectrum of quasiparticles in the range 
\begin{equation}
2 t' k_x^2  \lesssim |\Delta| 
\end{equation}
We see therefore that the gap opens up in the neighborhood 
of the saddle-points. The size of the part of the Fermi line  
destroyed is bounded by $\sqrt{|\Delta| / t'}$, in units
of the inverse lattice spacing.

The formation of the quasiparticle gap has its origin in the
hybridization of modes at different saddle-points, as a
consequence of the enhanced scattering with momentum transfer
exactly equal to ${\bf Q}$. Quite remarkably, this is an
effect that can be studied in the weak coupling regime of
the model, and the gap appears for arbitrarily small strength
$U$ of the interaction. From the technical point of view, the
discussion carried out in this section parallels the treatment
of the one-particle Green functions in the usual description of
the superconducting instability\cite{landau}. 
However, it is clear that the 
physical setting is quite different. In the present situation, 
the condensate is made of particle-hole pairs with a nonvanishing
average projection of the spin. The fact that a macroscopic
number of these pairs has been formed is what forces the
quasiparticles to live out of the range already excited by
the condensate.

An important issue concerns the spontaneous breakdown of the
spin-rotational symmetry in the condensate. Let us consider
the model at zero temperature regarding this matter. It
is clear that the nonvanishing average spin cannot
have in principle any preferred
direction in space. Recalling our definition in Eq. (\ref{gap}),
a real value of $\Delta $ implies that the spin of the
condensate points in the $x$ direction, since
\begin{equation}
   \int d^2 k d\omega \langle \Psi^{+}_{A\uparrow} ({\bf k})
 \Psi_{B\downarrow} ({\bf k}+{\bf Q}) \rangle  +
   \int d^2 k d\omega \langle \Psi^{+}_{B\downarrow} 
                            ({\bf k}+{\bf Q})
 \Psi_{A\uparrow} ({\bf k}) \rangle +
 A \leftrightarrow B   =   2 (\Delta + \Delta^{*})/U
\end{equation}
A purely imaginary value of $\Delta $ implies otherwise that
the spin of the condensate lies in the $y$ direction.
Finally, it may also be that the nonvanishing mean value is
realized for the $z$ component of the spin 
\begin{equation}
   \int d^2 k d\omega \langle \Psi^{+}_{A\uparrow} ({\bf k})
 \Psi_{B\uparrow} ({\bf k}+{\bf Q}) \rangle  -
   \int d^2 k d\omega \langle \Psi^{+}_{A\downarrow} ({\bf k})
 \Psi_{B\downarrow} ({\bf k}+{\bf Q}) \rangle +
 A \leftrightarrow B   \neq  0
\end{equation}

In the ground state of the model at zero temperature, the
spin of the condensate has to point in a definite direction
and the SU(2) rotational symmetry is spontaneously broken.
As a consequence, two Goldstone bosons arise in the spectrum,
which correspond to the spin waves that propagate on top of
the particle-hole condensate. These are the gapless excitations
of the model, together with the quasiparticle excitations that
exist sufficiently far away from the saddle-points.

\section{Superconducting instability}

We now turn to the instability that arises from the
divergent flow of Eqs. (\ref{iu1}) and (\ref{iu2}). The
integral of these equations depends on the 
position of the chemical potential with respect to the
VHS. For this reason, it is crucial to know how $\mu$
depends on the cutoff $\Lambda $ as this is progressively
lowered.

The issue of the renormalization of the chemical potential
has to be treated necessarily in the framework of the
wilsonian RG approach. As the high-energy modes are
integrated out at the scale $\Lambda $, $\mu $ shifts
its position by a quantity propotional to $d\Lambda $.
At the same time, it is the chemical potential
which sets the level to measure the energy cutoff,
as shown graphically in Fig. \ref{muren}.
The outcome is that $\mu $ adjusts itself at each step
of the RG process, until the point in which the cutoff
$\Lambda $ is lowered down to the final chemical potential.

At the computational level, the shift of $\mu $ is given
by the frequency and momentum-independent part of the
electron self-energy, with intermediate states taken from
the high-energy modes being integrated. The renormalization 
is proportional to the charge of the occupied states in 
the lower slice of width
$d\Lambda $, which couples through the forward-scattering
vertex $F$ in the usual Hartree and exchange diagrams.
The RG equation for the chemical potential reads
\begin{equation}
\frac{d\mu}{d\Lambda } =  F(\mu - \Lambda) \;  n(\mu - \Lambda) 
\label{diff}
\end{equation}
The perturbative approach is further improved by 
incorporating the renormalization of the $F$ vertex, 
which bears a well-known dependence on the energy 
scale measured from the VHS\cite{prl,charge}
\begin{equation}
F(\varepsilon ) \approx F_0/(1 - F_0 \log (|\varepsilon |)/
(4\pi^2 t))
\end{equation}

When the density of states $n(\varepsilon )$ is a smooth
function of the energy, the integration of high-energy modes
produces a steady downward flow of $\mu $. The physical
interpretation of this effect corresponds to the upward
displacement of the one-particle levels due to the repulsive
electronic interaction. In the neighborhood of the VHS,
the dynamics of $\mu $ becomes highly nonlinear given the
singular behavior of the density of states in Eq. (\ref{diff}).
It turns out that, in a certain range of initial values, the
chemical potential is renormalized down to the VHS and
precisely pinned to it in the low-energy regime.
As stated in the Introduction, this result pertains to
a statistical description in terms of the grand canonical 
ensemble. The physical picture is appropriate then for an
open system in contact with a charge reservoir, which sets 
the bare value $\mu_0 $ of the chemical potential.

\begin{figure}
\epsfxsize=7cm 
\centerline{\epsfbox{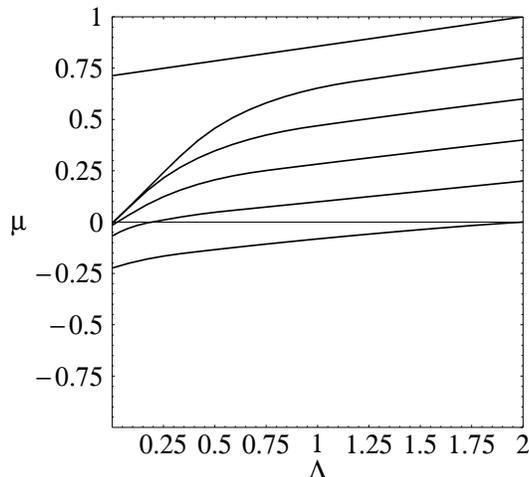}} 
\caption{Scaling of the chemical potential as a function of the
high-energy cutoff. The results correspond to the Hubbard 
coupling $U = 4t$.}
\label{mudiag}
\end{figure}

In order to evaluate the influence of the VHS on the
renormalization of the chemical potential, we have solved Eq.
(\ref{diff}) with the approximate density of states
\begin{eqnarray}
n(\varepsilon) & = &  c  \log (t/|\varepsilon|) /(4\pi^2 t)
 \;\;\;\;\;\;\;\; {\rm for } \;\;\;\;\;\;\;\; 
           \left|\varepsilon \right| \leq 0.5t          \\
  &   &    {\rm  const.}   \;\;\;\;\;\;\;\;\;\;\;\;\;\;\;\;
 \;\;\;\;\;\;\;\; {\rm for } \;\;\;\;\;\;\;\; 
          \left|\varepsilon \right|  >  0.5t
\end{eqnarray}
This expression has the correct normalization for the logarithmic
singularity in the 2D square lattice. The behavior of the 
integrals of Eq. (\ref{diff}) with such a density of states is
shown in Fig. \ref{mudiag}. It is manifest that, for initial
values of the chemical potential $\mu_0 \lesssim t$ above
the singularity, the final renormalized value of $\mu $ lies very
close to the VHS.
These results are important to assure that the enhancement
of the instabilities due to the divergent density of states does
not rely on fine-tuning the Fermi level to the VHS, as the 
chemical potential tends to pin itself in a natural way to the
singularity.

\begin{figure}
\epsfxsize=7cm 
\centerline{\epsfbox{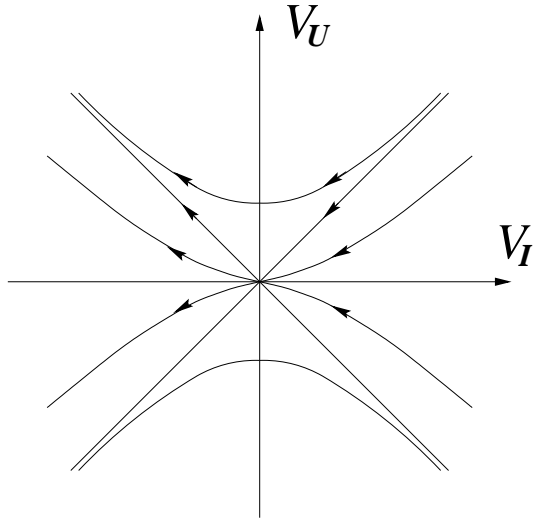}}
\caption{Flow of the renormalized BCS couplings in the  
$\left( V_I , V_U  \right)$ plane.}
\label{flowvv}
\end{figure}

The integrals of Eq. (\ref{diff}) can be used now to find the 
solutions of Eqs. (\ref{iu1}) and (\ref{iu2}) displaying the
superconducting instability.
The form of the flow in the coupling constant space is shown
in Fig. \ref{flowvv}. In the case of bare repulsive 
interactions, either the BCS couplings scale to zero
for $V_{I} > V_{U}$, or there is an
unstable flow giving rise to the superconducting instability
when $V_{I} < V_{U}$. The latter instance is realized in
lattice models which have a nearest-neighbor attractive 
interaction $V$
besides the on-site $U$ repulsive interaction. When
$V < 0$, the bare coupling $V_{I} = U + 4V$ is obviously
smaller than the bare coupling $V_{U} = U - 4V$. We are however
more interested in the case of the pure Hubbard
model, in which the bare couplings lie in the diagonal of the
first quadrant in Fig. \ref{flowvv}.

The couplings read directly from the hamiltonian of the Hubbard
model correspond to the boundary between the regions
of stable and unstable flow. This means that the slightest
perturbation may drive the system to either of the two sides,
which stresses the role played by the irrelevant operators under
these conditions. There are actually perturbations that fade away
when the theory is scaled to low energies, but that may be 
important because they may destabilize the flow in the BCS 
channel.

\begin{figure}
\epsfxsize=7cm 
\centerline{\epsfbox{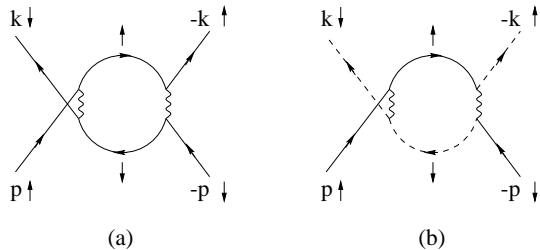}} 
\caption{Particle-hole corrections to the BCS vertices in the
Hubbard model.}
\label{iuwave}
\end{figure}

In the particular case of the Hubbard model, such irrelevant
perturbations are given by the iteration of
particle-hole diagrams of the type shown in Fig. \ref{iuwave}.
Apart from the particle-particle diagrams, these are the only
corrections that arise from the bare couplings of the model, and they
are not enhanced at low energies since the particle-hole bubbles
do not have the appropriate kinematics to be of order $\sim
d\Lambda $ in the wilsonian approach\cite{sh}.

The iteration of the bubbles in Fig. \ref{iuwave} gives rise to
antiscreening diagrams, i. e. to corrections that add to the
bare repulsive interaction. We recall that the particle-hole
bubble with total momentum about ${\bf Q}$ is enhanced with the 
factor $c'$ given after Eq. (\ref{chi1}), while that with momentum 
about the origin is proportional to the factor $c$ given
after Eq. (\ref{dos}). As long as in the present paper we
remain in the range $t' < 0.276 \; t$, we have that $c'$ is greater
than $c$, and we face the instance in which the irrelevant
perturbations make $V_{U}$ slightly larger than $V_{I}$
at the beginning of the RG flow.

We have solved the RG equations (\ref{iu1}) and (\ref{iu2})
taking as initial values for $V_I$ and $V_U$ the result of
adding the ladder series built from the diagrams in
Fig. \ref{iuwave}, with a bare Hubbard coupling $U = 4t$.
Moreover, in the resolution we have introduced
the dependence of $\mu $ on $\Lambda $ that arises from Eq.
(\ref{diff}). This is one of the main accomplishments of our
RG procedure, since the knowledge of how the VHS is approached is
essential to regularize the effect of the divergent density of
states.

\begin{figure}
\epsfxsize=7cm 
\centerline{\epsfbox{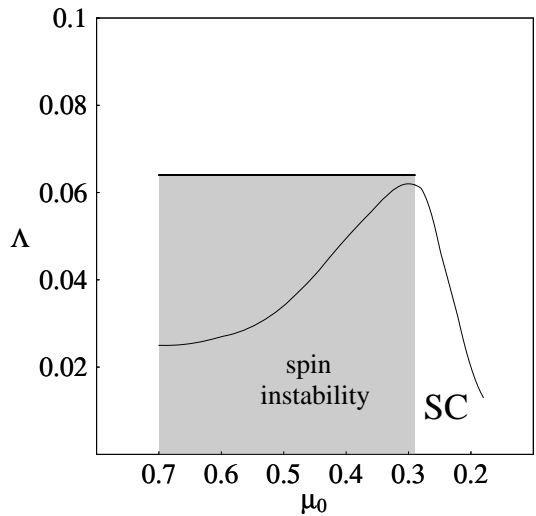}} 
\caption{Plot of the energy scale of the superconducting
instability (thin line) and of the transition to the
spin instability phase (shaded region).}
\label{filldiag}
\end{figure}

The results can be synthesized in the determination of the line at
which the transition to the superconducting state takes place in
the model. That is characterized by the energy at which the BCS
couplings grow large or, more conveniently, by the point at which
these couplings have a singularity. This depends on the initial 
position $\mu_0$ of the chemical potential, and it has been 
represented as a function of this variable in Fig. \ref{filldiag}.

We find that the BCS couplings diverge only for values
of $\mu_0 $ in the range of attraction to the VHS, that is when
the renormalized chemical potential is pinned to the
singularity. There is an optimal value of
$\mu_0 $ for which the scale of the transition reaches a maximum,
as the chemical potential stays closer to the VHS during a
greater part of the RG flow. For lower values of $\mu_0 $, the
scale of the instability decreases, as a consequence of the fact
that the renormalized chemical potential is not precisely pinned
then to the VHS.

We have also represented in Fig. \ref{filldiag} the energy at 
which the spin instability opens up, according to the estimate of
Section V. We realize that this scale is always above the
energy at which the singularity develops in the BCS channel,
whenever the spin instability exists in the system. This happens
for values of $\mu_0 $ higher than the optimal one. For lower 
values, the renormalized chemical potential deviates from the VHS 
by an amount even larger than the gap that would be due to the
spin instability, so that this does not find the conditions to
develop. We have then a picture in which the pairing instability 
exists alone for $\mu_0 $ below the optimal doping, but it is 
actually precluded above that level since the spin instability 
sets in before with the formation of a gap in the quasiparticle 
spectrum.

We comment finally on the symmetry of the condensate wavefunction.
The fact that the Umklapp interaction $V_U$ becomes increasingly 
repulsive when approaching the instability implies that the 
wavefunction must have opposite signs in the saddle-points $A$ and
$B$. As long as in the unstable flow we approach the asymptotic
regime $V_I = - V_U$, the response function for the $d$-wave
operator 
\begin{equation}
\Psi^{+}_{A\uparrow} ({\bf k}) \Psi^{+}_{A\downarrow} (-{\bf k})  -
 \Psi^{+}_{B\uparrow} ({\bf k}) \Psi^{+}_{B\downarrow} (-{\bf k}) + 
{\rm h. c.}
\end{equation}
develops a singularity at the frequency where the coupling
$V_I - V_U$ blows up. By the same token, it is easily seen that
the response function for the $s$-wave operator does not display
any divergence at low energies. Without the need of knowing 
precisely the shape of the gap, we may assure then that the 
symmetry of the order parameter is of $d$-wave type, with nodal
lines at the bisectors of the four quadrants. This is in 
agreement with the results of more general analyses, which 
show that the symmetry of the order parameter can be ascertain
from the topology of the Fermi line alone\cite{inst}.

\section{Conclusions}

In this paper we have presented a study of the different
phases of the system of electrons interacting near a Van
Hove singularity, when the correlations at momentum 
${\bf Q} \equiv (\pi , \pi )$ prevail over those at zero 
momentum. In the context of a model with nearest-neighbor 
and next-to-nearest-neighbor hopping, this happens for 
$0 < t' < 0.276 \; t$, 
according to the comparison of the prefactors $c$ and $c'$
that appear in Eqs. (\ref{dos}) and (\ref{chi1}), respectively.
We have applied a wilsonian RG approach following the same 
lines developed by Shankar in Ref \onlinecite{sh} for the
analysis of Fermi liquid theory. We have paid attention to
the spin degrees of freedom when considering the different
interactions, what has allowed us 
to discern the universality classes of the system.

We have seen that, regarding the spin correlations, there is
a universality class characterized by a spin instability in
the low-energy theory, in
opposition to the regime of couplings with smooth behavior of
the correlators for the spin operators. In the case of the
extended Hubbard model with on-site interaction $U$ and
nearest-neighbor interaction $V$, the spin instability arises
for $U > 0$, irrespective of the value of $V$, and it is
absent for $U < 0$.

Several authors have previously considered the competition 
between the spin and the superconducting instabilites
in the universality class corresponding to the divergent 
flow in the upper half-plane of
Fig. \ref{flowdd}\cite{dwave,iof,jpn,prl,ren,ren2,binz}.
Our analysis has shed light into a number of features 
of the spin instability. We have seen that this 
takes place through the condensation of particle-hole pairs with
momentum ${\bf Q}$. The fact that a macroscopic
number of these pairs has been formed is what forces the
quasiparticles to live out of the range $\Delta $ excited by   
the condensate in the neighborhood of the saddle-points. 
The Fermi line is destroyed 
in a region whose size is bounded by $\sqrt{|\Delta| / t'}$, in 
units of the inverse lattice spacing. 

This effect provides a
paradigm for the disappearance of the Fermi surface of an electron
system which differs from the understanding of such a phenomenon
in Mott-Hubbard insulators. Those systems are supposed to be in a
strong-coupling regime, in which the double occupancy of each 
lattice site is highly suppressed. In our case,
we need otherwise to constrain the Fermi level near the VHS, 
departing sensibly from half-filling as $t'$ is increased. 
Most remarkably, the instability takes place no matter how small
the bare couplings may be in the above picture.
This is what ultimately allows
to discern the symmetry breaking in the ground state within our RG 
approach.

We have seen that two different behaviors arise also
in the space of couplings for the BCS channel, starting from
bare repulsive interactions $V_I$ and $V_U$.
The $d$-wave superconducting
instability develops in models corresponding to the region with
unstable flow in the upper half-plane of Fig. \ref{flowvv}. 
This is the case of the
extended Hubbard model with $U > 0$ and attractive interaction $V$. 
When $U > 0$ and the nearest-neighbor interaction is repulsive,
the couplings in the BCS channel scale down to zero.
The Hubbard model with just on-site interaction is placed at first
sight on the boundary between the regions with stable and unstable
behavior. We have shown that the model has irrelevant
perturbations that drive the system towards the side with divergent
RG flow. Since the departure from the limit behavior is weak, the
superconducting instability is overshadowed by the spin
instability, up to a point of optimal doping beyond which the latter
is absent. 

The use of the wilsonian RG approach provides some advantages over
other RG methods, the most important being the possibility of
studying the renormalization of the chemical potential. Given the
divergent behavior of the density of states at the VHS, it is
clear that all the positions of the Fermi level cannot be equally
stable. The scaling of the chemical potential can be obtained by
letting it free to evolve and computing the shift from the 
integration of high-energy modes near the cutoff at each RG
step. Following this procedure, we have seen that there is a range
of attraction near the VHS where the chemical potential is
renormalized down to the singularity. This guarantees the
naturalness of the different instabilities since, rather than
relying on the fine-tuning of the Fermi level, they arise from its
precise pinning to the VHS in the low-energy effective theory.

\section*{Appendix}

In this section we compute the imaginary part of some of
the susceptibilities of the model. It turns out that the
particle-hole susceptibility $\chi_{ph} ({\bf Q},\omega )$
and the particle-particle susceptibility $\chi_{pp}
({\bf 0},\omega )$ have a nontrivial imaginary part, while this
vanishes for $\chi_{ph} ({\bf 0},\omega )$ and $\chi_{pp}
({\bf Q},\omega )$ at any finite frequency.

In our model, the susceptibility $\chi_{ph} ({\bf Q}, \omega )$
is given by
\begin{equation}
\chi_{ph} ({\bf Q},\omega ) =  i \int \frac{d \omega_q}{2\pi }
    \int \frac{d^2 q}{(2\pi )^2}
  \frac{1}{\omega + \omega_q - \varepsilon_A ({\bf q}) 
     + i \epsilon \; {\rm sgn}  (\omega + \omega_q )  } 
  \frac{1}{\omega_q - \varepsilon_B ({\bf q}) 
     + i \epsilon \; {\rm sgn}  (\omega_q )  } 
\label{susc}
\end{equation}
where the energy cutoff is implicit in the integration over
the momenta. According to the standard prescription, the
imaginary part of (\ref{susc}) is given by
\begin{equation}
{\rm Im} \; \chi_{ph} ({\bf Q},\omega ) =
    - 2 \pi^2  \int \frac{d \omega_q}{2\pi }
     \int \frac{d^2 q}{(2\pi )^2}
     {\rm sgn}  (\omega + \omega_q )  \; {\rm sgn}  (\omega_q ) \;
  \delta ( \omega + \omega_q - \varepsilon_A ({\bf q}) )  \;
  \delta ( \omega_q - \varepsilon_B ({\bf q}) )
\label{suscp}
\end{equation}
In the limit of small $t'$, Eq. (\ref{suscp}) leads to a
quantity which does not depend on the frequency.
Taking $\omega > 0$, we have
\begin{eqnarray}
{\rm Im} \; \chi_{ph} ({\bf Q},\omega ) & = &
        \frac{1}{4\pi } \int d^2 q  \;
      \delta (\omega + 2t(q_x^2 - q_y^2))           \\
  & = &  \frac{1}{8\pi t} \int_{-q_0}^{q_0} d q_x
      \frac{1}{ \sqrt{q_x^2 + \omega /(2t) }  }
\end{eqnarray}
where $q_0 = \sqrt{ \frac{\omega (t-2t')}{8tt'} }$.
After a little of algebra, we obtain
\begin{eqnarray}
{\rm Im} \; \chi_{ph} ({\bf Q},\omega ) & = &
    \frac{1}{8\pi t}
 \log \left(\frac{t}{2t'} + \sqrt{\frac{t^2}{4t'^2} - 1}\right) \\
  & = & c'/(8 \pi t)
\end{eqnarray}

We see therefore that the imaginary part is equal to $\pi /2$
times the prefactor of $\log (\Lambda )$ in the real part of the
susceptibility. It can be checked that
the same relation holds between the real and the imaginary part
of the particle-particle susceptibility $\chi_{pp}
({\bf 0},\omega )$.

Turning now to the susceptibility $\chi_{ph} ({\bf 0},\omega )$,
we have
\begin{eqnarray}
{\rm Im} \; \chi_{ph} ({\bf 0},\omega ) & = &
   {\rm Re } \;  \int \frac{d \omega_q}{2\pi }
         \int \frac{d^2 q}{(2\pi )^2}
  \frac{1}{\omega + \omega_q - \varepsilon_A ({\bf q}) 
     + i \epsilon \; {\rm sgn}  (\omega + \omega_q )  } 
  \frac{1}{\omega_q - \varepsilon_A ({\bf q}) 
     + i \epsilon \; {\rm sgn}  (\omega_q )  }          \\
     & = & - 2 \pi^2 \int \frac{d \omega_q}{2\pi }
         \int \frac{d^2 q}{(2\pi )^2}
     {\rm sgn}  (\omega + \omega_q )  \; {\rm sgn}  (\omega_q ) \;
  \delta ( \omega + \omega_q - \varepsilon_A ({\bf q}) )  \;
  \delta ( \omega_q - \varepsilon_A ({\bf q}) )         \\
    & = & - \frac{1}{4\pi } \delta (\omega ) \int d^2 q
\label{sferro}
\end{eqnarray}
We see then that the imaginary part of the susceptibility is 
zero for any finite value of the frequency.

A result similar to (\ref{sferro}) is obtained for the imaginary
part of the susceptibility $\chi_{pp} ({\bf Q},\omega )$. In
this channel, the pole that arises after summing up
the ladder series corresponds to the appearance of excited states
in the spectrum. We conclude therefore that the breakdown of 
symmetry through a mechanism of condensation can only take place in 
the particle-particle channel at zero momentum and in the 
particle-hole channel at momentum ${\bf Q}$, as stated in the 
text.


\begin{references}


\bibitem{dago}
See, for instance, E. Dagotto, Rev. Mod. Phys. {\bf 66}, 763 
(1994), and 
P. W. Anderson, {\em The Theory of Superconductivity in the 
High-$T_c$ Cuprates} (Princeton Univ., Princeton, 1997).

\bibitem{sh}
R. Shankar, Rev. Mod. Phys. {\bf 66}, 129 (1994).

\bibitem{pol}
J. Polchinski
in {\em Proceedings of the 1992 TASI in Elementary Particle
Physics}, J. Harvey and J. Polchinski eds. (World Scientific,
Singapore, 1992). 

\bibitem{ex}
W. Metzner, C. Castellani and C. di Castro, Adv. Phys. {\bf 47},
3 (1998).

\bibitem{bares}
P.-A. Bares and X.-G. Wen, Phys. Rev. B {\bf 48}, 8636 (1993).

\bibitem{wo}
J. Gan and E. Wong, Phys. Rev. Lett. {\bf 71}, 4226 (1993).

\bibitem{sh2}
A. Houghton, H.-J. Kwon, J. B. Marston and R. Shankar, J. Phys.
Condens. Matter {\bf 6}, 4909 (1994).

\bibitem{it}
C. Castellani and C. Di Castro, Physica C {\bf 235-240}, 99 (1994).
C. Castellani, C. Di Castro and A. Maccarone, Phys. Rev. B {\bf 55},
2676 (1997).
C. Castellani, S. Caprara, C. Di Castro and A. Maccarone, 
Nucl. Phys. B {\bf 594}, 747 (2001).

\bibitem{wi}
C. Nayak and F. Wilczek, Nucl. Phys. B {\bf 417}, 359 (1994).

\bibitem{us}
J. Gonz\'alez, F. Guinea and M. A. H. Vozmediano,
Nucl. Phys. B {\bf 424}, 595 (1994).

\bibitem{ch}
S. Chakravarty, R. E. Norton and O. F. Syljuasen, Phys. Rev.
Lett. {\bf 74}, 1423 (1995).

\bibitem{inst}
J. Gonz\'alez, F. Guinea and M. A. H. Vozmediano,
Phys. Rev. Lett. {\bf 79}, 3514 (1997).

\bibitem{early}
J. Labb\'e and J. Bok, Europhys. Lett. {\bf 3}, 1225 (1987).
J. Friedel, J. Phys. (Paris) {\bf 48}, 1787 (1987); {\bf 49},
1435 (1988).
R. S. Markiewicz and B. G. Giessen, Physica
{\bf 160C}, 497 (1989).
C. C. Tsuei {\em et al.}, Phys. Rev. Lett. {\bf 65},
2724 (1990).
D. M. Newns {\em et al.}, Phys. Rev. Lett. {\bf 69},
1264 (1992).

\bibitem{rev}
A review of the Van Hove scenario for high-$T_c$
superconductivity has been made by
R. S. Markiewicz, J. Phys. Chem. Sol. {\bf 58}, 1179
(1997).

\bibitem{dwave}
H. J. Schulz, Europhys. Lett. {\bf 4}, 609 (1987).
P. Lederer, G. Montambaux and D. Poilblanc, J. Phys.
(Paris) {\bf 48}, 1613 (1987).
J. E. Dzyaloshinskii, Pis'ma Zh. Eksp. Teor. Fiz. {\bf
46}, 97 (1987) [JETP Lett. {\bf 46}, 118 (1987)].

\bibitem{pin}
J. Gonz\'alez, F. Guinea and M. A. H. Vozmediano, Europhys.
Lett. {\bf 34}, 711 (1996); report cond-mat/9502095.

\bibitem{iof}
L. B. Ioffe and A. J. Millis, Phys. Rev. B {\bf 54}, 3645
(1996).

\bibitem{kohn}
D. Zanchi and H. J. Schulz, Phys. Rev. B {\bf 54}, 9509
(1996).

\bibitem{liu}
D. Z. Liu and K. Levin, Physica {\bf 275C}, 81 (1997).

\bibitem{pat}
P. C. Pattnaik {\em et al.}, Phys. Rev. B {\bf 45}, 5714
(1992).

\bibitem{nucl}
J. Gonz\'alez, F. Guinea and M. A. H. Vozmediano, Nucl.
Phys. B {\bf 485}, 694 (1997).

\bibitem{men}
D. Menashe and B. Laikhtman, Phys. Rev. B {\bf 59}, 13592
(1999).

\bibitem{george}
G. Kastrinakis, Physica C {\bf 340}, 119 (2000).

\bibitem{kat}
V. Yu. Irkhin and A. A. Katanin, Phys. Rev. B {\bf 64}, 205105
(2001).

\bibitem{jpn}
J. V. Alvarez, J. Gonz\'alez, F. Guinea and M. A. H. Vozmediano,
J. Phys. Soc. Jpn. {\bf 67}, 1868 (1998).

\bibitem{prl}
J. Gonz\'alez, F. Guinea and M. A. H. Vozmediano,
Phys. Rev. Lett. {\bf 84}, 4930 (2000).

\bibitem{kl}
W. Kohn and J. M. Luttinger, Phys. Rev. Lett.
{\bf 15}, 524 (1965).

\bibitem{chu}
For a review, see M. A. Baranov, A. V. Chubukov 
and M. Yu. Kagan, Int. J. Mod. Phys. B {\bf 6}, 2471 (1992).

\bibitem{ren}
C. J. Halboth and W. Metzner, Phys. Rev. B {\bf 61}, 7364
(2000); Phys. Rev. Lett. {\bf 85}, 5162 (2000).

\bibitem{ren2}
C. Honerkamp, M. Salmhofer, N. Furukawa and T. M. Rice,
Phys. Rev. B {\bf 63}, 35109 (2001).

\bibitem{binz}
B. Binz, D. Baeriswyl and B. Dou\c{c}ot,
Eur. Phys. J. B {\bf 25}, 69 (2002).

\bibitem{mark}
R. S. Markiewicz, J. Phys.: Condens. Matter {\bf 2}, 665
(1990).

\bibitem{newns}
D. M. Newns, P. C. Pattnaik and and C. C. Tsuei, 
Phys. Rev. B {\bf 43}, 3075 (1991).

\bibitem{kat2}
V. Yu. Irkhin, A. A. Katanin and M. I. Katsnelson, 
Phys. Rev. B {\bf 64}, 165107 (2001).

\bibitem{sa}
C. Honerkamp and M. Salmhofer, Phys. Rev. Lett. {\bf 87},
187004 (2001).

\bibitem{sor}
S. Sorella, R. Hlubina and F. Guinea, Phys. Rev. Lett.
{\bf 78}, 1343 (1997).

\bibitem{charge}
J. Gonz\'alez, Phys. Rev. B {\bf 63}, 45114 (2001).

\bibitem{lh}
H. Q. Lin and J. E. Hirsch, Phys. Rev. B {\bf 35}, 3359 (1987).

\bibitem{lutt1}
J. S\'olyom, Adv. Phys. {\bf 28}, 201 (1979).

\bibitem{sch}  
H. J. Schulz, in {\em Correlated Electron Systems}, Vol. 9,
ed. V. J. Emery (World Scientific, Singapore, 1993).

\bibitem{agd}
A. A. Abrikosov, L. P. Gorkov and I. E. Dzyaloshinski,
{\em Methods of Quantum Field Theory in Statistical Physics},
Chap. 7 (Dover, New York, 1975).

\bibitem{landau}
E. M. Lifshitz and L. P. Pitaevskii, {\em Statistical Physics, 
Part 2}, Chap. 5 (Pergamon Press, Oxford, 1980).







\end{references}
\end{document}